\documentclass[sigconf,natbib=true,screen=true]{acmart}

\usepackage{bm}
\usepackage{caption}
\usepackage{graphicx}
\usepackage{subfigure}
\usepackage{amsmath}
\usepackage{booktabs}
\usepackage{threeparttable}
\usepackage{multirow}
\usepackage{enumitem}
\usepackage{xcolor}
\usepackage{mathtools}
\usepackage[ruled]{algorithm2e}
\usepackage{setspace}

\AtBeginDocument{%
  }

\copyrightyear{2023}
\acmYear{2023}
\setcopyright{rightsretained}
\acmConference[CIKM '23]{Proceedings of the 32nd ACM International Conference on Information and Knowledge Management}{October 21--25, 2023}{Birmingham, United Kingdom}
\acmBooktitle{Proceedings of the 32nd ACM International Conference on Information and Knowledge Management (CIKM '23), October 21--25, 2023, Birmingham, United Kingdom}
\acmDOI{10.1145/3583780.3614947}
\acmISBN{979-8-4007-0124-5/23/10}

\begin{document}

\title{L$^2$R: Lifelong Learning for First-stage Retrieval with Backward-Compatible Representations}

\author{Yinqiong Cai}
\affiliation{
 \institution{CAS Key Lab of Network Data Science and Technology, ICT, CAS}
 \institution{University of Chinese Academy of Sciences}
 \city{Beijing}
 \country{China}
}
\email{caiyinqiong18s@ict.ac.cn}

\author{Keping Bi}
\affiliation{
 \institution{CAS Key Lab of Network Data Science and Technology, ICT, CAS}
 \institution{University of Chinese Academy of Sciences}
 \city{Beijing}
 \country{China}
}
\email{bikeping@ict.ac.cn}

\author{Yixing Fan}
\affiliation{
 \institution{CAS Key Lab of Network Data Science and Technology, ICT, CAS}
 \institution{University of Chinese Academy of Sciences}
 \city{Beijing}
 \country{China}
}
\email{fanyixing@ict.ac.cn}

\author{Jiafeng Guo}
\authornote{Jiafeng Guo is the corresponding author.}
\affiliation{
 \institution{CAS Key Lab of Network Data Science and Technology, ICT, CAS}
 \institution{University of Chinese Academy of Sciences}
 \city{Beijing}
 \country{China}
}
\email{guojiafeng@ict.ac.cn}

\author{Wei Chen}
\affiliation{
 \institution{CAS Key Lab of Network Data Science and Technology, ICT, CAS}
 \institution{University of Chinese Academy of Sciences}
 \city{Beijing}
 \country{China}
}
\email{chenwei2022@ict.ac.cn}

\author{Xueqi Cheng}
\affiliation{
 \institution{CAS Key Lab of Network Data Science and Technology, ICT, CAS}
 \institution{University of Chinese Academy of Sciences}
 \city{Beijing}
 \country{China}
}
\email{cxq@ict.ac.cn}
\renewcommand{\shortauthors}{Yinqiong Cai et al.}
\newcommand{\modelname}[1]{L$^2$R}

\begin{abstract}
First-stage retrieval is a critical task that aims to retrieve relevant document candidates from a large-scale collection.
While existing retrieval models have achieved impressive performance, they are mostly studied on static data sets, ignoring that in the real-world, the data on the Web is continuously growing with potential distribution drift.
Consequently, retrievers trained on static old data may not suit new-coming data well and inevitably produce sub-optimal results.
In this work, we study lifelong learning for first-stage retrieval, especially focusing on the setting where the emerging documents are unlabeled since relevance annotation is expensive and may not keep up with data emergence.
Under this setting, we aim to develop model updating with two goals: (1) to effectively adapt to the evolving distribution with the unlabeled new-coming data, and (2) to avoid re-inferring all embeddings of old documents to efficiently update the index each time the model is updated. 

We first formalize the task and then propose a novel Lifelong Learning method for the first-stage Retrieval, namely \modelname{}.
\modelname{} adopts the typical memory mechanism for lifelong learning, and incorporates two crucial components: (1) selecting \textit{diverse support negatives} for model training and memory updating for effective  model adaptation, 
and (2) a \textit{ranking alignment objective} to ensure the backward-compatibility of representations to save the cost of index rebuilding without hurting the model performance.
For evaluation, we construct two new benchmarks from LoTTE and Multi-CPR datasets to simulate the document distribution drift in realistic retrieval scenarios.
Extensive experiments show that \modelname{} significantly outperforms competitive lifelong learning baselines.
\end{abstract}



\maketitle

\vspace{-2mm}
\section{Introduction}
First-stage retrieval aims to quickly retrieve a few relevant document candidates from a large-scale collection, which has become a core component in information retrieval (IR) applications~\cite{guo2022semantic, zhao2022dense}.
While retrieval models based on pre-trained language models (PLMs)~\cite{xiong2020approximate,santhanam2021colbertv2,zhou2022master,liu2022retromae} have demonstrated impressive performance, most of them are studied on  static datasets, neglecting that in the real world, new documents are continuously emerging on the Web.
For example, when a new event (e.g., ChatGPT) breaks out, a large number of documents on this topic were generated and shared, accompanied by booming information needs regarding the topic (searching for not only new documents but also old ones).
The emerging documents and queries on the new topics may cause the distribution of retrieval collection to drift over time. 
Consequently, directly applying the model trained on previous data to the new collection is obviously not an optimal solution.
Then, how can we continuously learn a retrieval model to adapt to the evolving data distribution effectively and efficiently? To study this problem, we formalize the task of lifelong learning for first-stage retrieval.

Lifelong learning ~\cite{chen2018lifelong, van2019three} has been widely studied in the machine learning community, especially on computer vision (CV) tasks~\cite{mai2022online, wang2023comprehensive}.
In a typical setting of lifelong learning, a model is set to learn with non-identically and independently distributed (non-I.I.D.) new-coming data~\cite{wang2023comprehensive}, with the goal of preserving acquired knowledge and learning new knowledge.
Most research~\cite{aljundi2019mir, li2017lwf, rusu2016pnn} in this field focuses on addressing the catastrophic forgetting issue~\cite{goodfellow2013empirical,mccloskey1989catastrophic}, i.e., the model's inability to perform well on previously seen data after being updated with new data. 
One representative paradigm for lifelong learning is the memory-based method~\cite{aljundi2019mir,chaudhry2019er}, which stores and replays historical samples while training on new data to mitigate the forgetting of acquired knowledge. 
These lifelong learning methods have been shown to be effective in various CV tasks~\cite{yoon2021ocs, ramakrishnan2020relationship}. 
However, there has been limited research on the lifelong learning problem for IR tasks.

In this paper, we study the task of lifelong learning for first-stage retrieval in a setting where new documents are unlabeled. We focus on this setting in our initial attempt because relevance annotation on the new data is expensive and may not catch up with data emergence. 
In this setting, besides the essential goal of general lifelong learning, i.e., preserving  acquired knowledge and learning new knowledge, it poses several new challenges: 
\begin{itemize}[leftmargin=1.2em,topsep=0pt,parsep=0pt]  
\item [1)] Without labeled positive samples, new data could have limited benefit to supervise model learning. Moreover, the unlabeled positives in new data could mislead the model if we simply take all new documents as irrelevant for training. 
\item [2)] It incurs significant costs to re-compute all document representations and rebuild the entire index each time the model is updated. It would be ideal to avoid repeated representation computation without harming model performance. 
\item [3)] The pairwise modeling of query-document pairs in IR makes the task more complicated, compared to the pointwise modeling of the classification tasks in CV. For any query, either seen or unseen, the model needs to achieve good retrieval performance on both new and old documents. 
\end{itemize}
Due to these challenges, existing lifelong learning methods in other fields cannot be directly applied to the retrieval task.
Although some work has explored the catastrophic forgetting issue of re-ranking models under the lifelong learning setting, no feasible solutions are proposed to solve it~\cite{lovon2021studying, gerald2022continual}.

To address the above challenges, we propose a memory-based Lifelong Learning method for first-stage Retrieval, named as \modelname{}. 
\modelname{} maintains a buffer to store the historical support negatives (i.e., negative samples that are important for learning the decision boundary of the model) for each training query, and when a session of unlabeled new documents arrives, it updates the model as follows:
1) To adapt the model to the new distribution, \modelname{} selects diverse support negatives from the unlabeled new data for model training, by estimating their confidence of being hard negatives and redundancy with other selected ones.
2) To balance the model's ability to preserve acquired knowledge and learn new knowledge, \modelname{} selects historical support documents distinct from the selected new samples and uses them together for model updating. 
3) To avoid re-inferring embeddings of old documents each time the model is updated, \modelname{} incorporates a novel ranking alignment objective to ensure the backward compatibility of document representations without harming retrieval performance.
Overall, through the selection strategy of \textit{diverse support negatives} and the \textit{ranking alignment objective} for compatible learning, \modelname{} enables effective and efficient retrieval model lifelong learning.

For evaluation, we construct two benchmarks based on the LoTTE~\cite{santhanam2021colbertv2} and Multi-CPR~\cite{long2022multi} datasets, namely LL-LoTTE and LL-MultiCPR, to simulate the realistic retrieval scenario where documents emerge continuously with distribution drift.
The empirical results on both benchmarks show that \modelname{} outperforms representative and state-of-the-art lifelong learning baselines in terms of metrics on both learning new data and addressing the forgetting issue. 
Moreover, our proposed ranking alignment objective achieves not only representation backward compatibility but also remarkably even better performance. 
We further confirm the advantages of our model through in-depth studies on the data selection strategy and the backward-compatible alignment objectives.

\vspace*{-2mm}
\section{Related work}

\vspace*{-0.5mm}
\noindent\textbf{Lifelong Learning.}
Lifelong learning~\cite{chen2018lifelong}, also referred to as continual learning~\cite{van2019three} or incremental learning~\cite{chaudhry2018riemannian}, has received much attention in building adaptive systems that are able to gain, retain, and transfer knowledge when facing non-stationary data streams.
Research in this field mainly focuses on solving the catastrophic forgetting issue~\cite{mai2022online, wang2023comprehensive}. There are three main method paradigms, including regularization-based~\cite{li2017lwf, kirkpatrick2017ewc}, architecture-based~\cite{rusu2016pnn, collier2020routing} and memory-based methods~\cite{rebuffi2017icarl, chaudhry2019er, aljundi2019mir,knoblauch2020optimal}.

Lifelong learning has been widely studied in various machine learning tasks~\cite{rebuffi2017icarl, yoon2021ocs, ramakrishnan2020relationship, shmelkov2017incremental}. 
Recently, ~\citet{mai2022online} surveyed a wide range of methods to address the lifelong learning problem for image classification.
In natural language processing tasks, the research on lifelong learning mostly focuses on pre-training~\cite{biesialska2020continual, wu2021pretrained}. For example, ~\citet{qin2022elle} proposed ELLE for incremental pre-training on emerging data efficiently. ~\citet{wu2021pretrained} compared the performance over the combination of five PLMs and four lifelong learning approaches.
However, to our best knowledge, there have been no studies on lifelong learning  for first-stage retrieval.

\noindent\textbf{First-stage Retrieval.}
In recent years, substantial efforts have been made on various retrieval models~\cite{guo2022semantic, zhao2022dense}, including both classical term-based methods like BM25~\cite{robertson2009probabilistic}, and more recent PLMs-based dense retrieval models~\cite{fan2022pre,lin2021pretrained, Ma, liu2023robustness}. 
PLMs-based retrieval models have compelling performance and are widely adopted in the industry. 
However, most existing studies are on static document sets, ignoring the realistic scenario wherein new documents continually arrive at the system.

Lifelong learning for information retrieval (IR) is an important but less-explored topic, including both the first-stage retrieval and re-ranking stage.
Recently, \citet{lovon2021studying} and \citet{gerald2022continual} studied the lifelong learning problem for PLMs-based re-ranking models. They observed the catastrophic forgetting issue in lifelong IR model learning. 
Later, \citet{mehta2022dsi++} studied continual learning for generative IR models~\cite{tay2022transformer}, in which they studied how to incrementally index new documents into the model parameters, instead of the distribution shift caused by newly emerged data.
Lifelong learning has been studied in image retrieval~\cite{shen2020towards, wan2022continual}.
However, the experiments were conducted on fine-grained image classification datasets, the settings of which completely differ from the realistic scenario for document retrieval.

\noindent\textbf{Compatible Representation Learning.}
Learning compatible representations~\cite{shen2020towards,ramanujan2022forward,hu2022learning,duggal2021compatibility} is a practical need in many scenarios, with the goal of ensuring the embeddings generated by different models are compatible.
For example, BCT~\cite{shen2020towards} and LCE~\cite{meng2021learning} learn compatible representations for image recognition, where the embeddings computed by the updated model are directly comparable to those generated by  previous models.
Specifically, BCT~\cite{shen2020towards} constrains the feature space by simultaneously enabling gradient flow from both the old and new classifiers. However, this method is not suitable for first-stage retrieval, since the relevance score is calculated on the embeddings directly and there are no classification layers.
LCE~\cite{meng2021learning} bridges the multiple feature spaces via a lightweight transformation function. However, they still need to re-compute all embeddings of the previous images, which is inefficient for large collection in retrieval.
Beyond these, representation compatibility has received increasing attention in asymmetric retrieval~\cite{duggal2021compatibility}, where the query and document use different models due to the constrained resources of the computing platform.
In contrast to these methods, we study compatible representation learning under the lifelong learning setting for first-stage retrieval.

\vspace*{-3mm}
\section{Task Description}

\vspace{-0.5mm}
\noindent\textbf{First-stage Retrieval.}
Given a query $q$ and a document collection $\mathcal{D}_0$, first-stage retrieval aims to find potentially relevant documents.
With a labeled training dataset $\smash{\mathcal{C}_0\!=\!\{(q, D_q^+)\}}$, where $q$ is a query and $\smash{d_q^+ \!\in\! D_q^+}$ is one of the relevant documents for $q$, we can build an initial retrieval model $\smash{f_0}$ using a dual-encoder architecture with a standard contrastive learning objective~\cite{guo2022semantic, zhao2022dense}. 
Then, the embeddings of documents in $\smash{\mathcal{D}_0}$ are extracted and indexed, and the retrieval is performed by estimating the similarity between the query embedding with the document embeddings in the index.

\noindent\textbf{Lifelong Learning for First-stage Retrieval.}
A stream of document sets $\{\smash{\mathcal{D}_1}, \cdots, \smash{\mathcal{D}_T}\}$ having different distributions arrive in $T$ sessions sequentially. 
Note that these new documents have no relevance labels.
For any session $t \!\in\! \{1,\cdots,T\}$, the lifelong learning algorithm $\mathcal{A}$ utilizes documents in $\smash{\mathcal{D}_t}$ to update $f_{t\!-\!1}$ to $f_{t}$, aiming to adapt the retriever to the new distribution,
\begin{equation}
\small
\label{eq:incremental_train}
\setlength{\abovedisplayskip}{0pt}
\setlength{\belowdisplayskip}{0pt}
\mathcal{A}_{t}:\left\langle f_{t\!-\!1}, \mathcal{C}_0, \mathcal{D}_t, M_{t-1}\right\rangle \rightarrow\left\langle f_{t}, M_{t}\right\rangle,
\end{equation}
where $\smash{M_{t\!-\!1}}$ and $\smash{M_t}$ are the external memory for session $t\!-\!1$ and $t$ respectively, which store useful information for lifelong model learning, e.g., a subset of training samples or historical versions of the model.
If the model updating is representation backward-compatible, at session $t$,  we only need to compute document embeddings for $\smash{\mathcal{D}_t}$ using model $\smash{f_{t}}$. The embeddings of $\smash{\mathcal{D}_{0:t\!-\!1}=\bigcup_{i=0}^{t-1} \mathcal{D}_{i}}$ that are computed with historical models can be reused when updating the index with existing techniques~\cite{johnson2019billion}.
Otherwise, we need to use $f_t$ to compute the embeddings for all documents in $\smash{\mathcal{D}_{0:t}}$ to rebuild the retrieval index.

\begin{figure*}[!t]
\setlength{\abovecaptionskip}{0pt}
\includegraphics[scale=0.45]{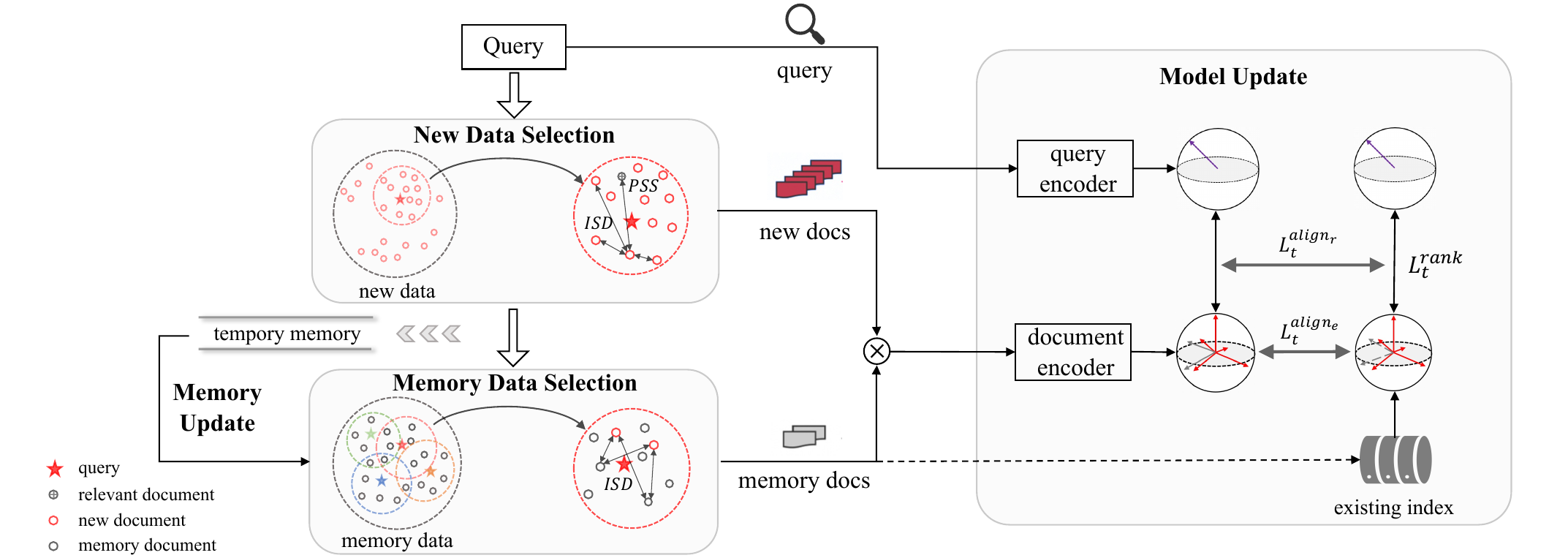}
\caption{Memory-based lifelong learning method for first-stage retrieval (\modelname{}).}
\label{fig:model}
\vspace{-5mm}
\end{figure*}

\vspace*{-3mm}
\section{Methodology}

\subsection{Overview of the Approach}
We employ the typical memory mechanism~\cite{chaudhry2019er,shim2021aser} in \modelname{} and maintain a restricted external memory to store a subset of historical documents for each training query. This memory mechanism enables the model to efficiently determine the replay samples to address the catastrophic forgetting issue, without browsing from the entire collection.
Based on the memory mechanism, for effective and efficient lifelong learning of retrieval models, \modelname{} incorporates two important components, including the selection strategy of \textbf{diverse support negatives}  and the \textbf{ranking alignment objective} for backward-compatible representation learning.

As shown in Algorithm~\ref{algo:overview}, when the newly emerged data $\smash{\mathcal{D}_t}$ arrives at session $t$, \modelname{} selects diverse support negative documents from the new data and the memory buffer $\smash{M_{t\!-\!1}}$ respectively, then updates the retriever from $\smash{f_{t-1}}$ to $\smash{f_t}$ with the selected samples, and finally updates the memory with the new data.
Next, we introduce the detailed data selection method (Section~\ref{sec:data_selection}), and the optimization objective for compatible learning (Section~\ref{sec:compatible_representation_learning}).

\begin{algorithm}[!t]
\setstretch{1.0}
\small
\caption{{\sc Overview of \modelname{}.}}
\label{algo:overview}
\LinesNumbered
\KwIn{Dataset $\mathcal{C}_0$, Retrieval model $f_0$, Memory buffer $M_0$ with $n$ slots for each query, Total sessions $T$}
\KwOut{Retrieval Model $f_{T}$}
    \For{$t \in \{1,\cdots,T\}$}{
        $f_t \leftarrow f_{t-1}$; $M_c \leftarrow \{\}$\;
        \For{$(q, d_q^+) \sim \mathcal{C}_0$}{
            $D_{\mathit{q}}^{\mathit{new}} \leftarrow \textit{NewDataSelection}(q, d_q^+, \mathcal{D}_t)$\;
            $M_c \leftarrow M_c \cup D_{\mathit{q}}^{\mathit{new}}$\;
            $\smash{D_{\mathit{q}}^{\mathit{mem}}} \leftarrow \textit{MemoryDataSelection}(q, \smash{D_{\mathit{q}}^{\mathit{new}}}, M_{t-1})$\;
            $f_t \leftarrow \textit{ModelUpdate}(q, \smash{d_q^+}, \smash{D_{\mathit{q}}^{\mathit{new}} \cup D_{\mathit{q}}^{\mathit{mem}}}, f_t)$\;
        }
        $\smash{M_t \leftarrow \textit{MemoryUpdate}\left(q, M_c, M_{t-1}, n\right)}$\;
    }
    \Return{$f_T$}
\end{algorithm}

\vspace{-3mm}
\subsection{Diverse Support Negative Selection} \label{sec:data_selection}
To effectively adapt the retriever to the new distribution, we desire to select support and diverse negatives for model training.
Thus, we define \textit{positive sample superiority} ($PSS$) and \textit{inter sample diversity} ($ISD$) criteria to instruct the data selection in each step.

Let $\bm{q}$ and $\bm{d}$ denote the embedding of query $q$ and document $d$, and $\smash{\bm{d_{_\parallel q}}}$ and $\smash{\bm{d_{\perp q}}}$ denote the projection of $\bm{d}$ on the directions that are horizontal and vertical to $\bm{q}$. Intuitively, $\smash{\bm{d_{_\parallel q}}}$ and $\smash{\bm{d_{\perp q}}}$ represent the information in $d$ that is related and unrelated to $q$ respectively.

\noindent \textbf{Definition 1 (Positive Sample Superiority).} 
The \textit{positive sample superiority} between $d$ and $\smash{d_q^+}$ for the query $q$ is given by
\begin{equation}
\small
\setlength{\abovedisplayskip}{1pt}
\setlength{\belowdisplayskip}{1pt}
PSS(d, d_q^+; q) = sign(\bm{d^{+}_{_\parallel q}} \!-\! \bm{d_{_\parallel q}}) \cdot \left\lVert\bm{d_{_\parallel q}^{+}} \!-\! \bm{d_{_\parallel q}}\right\rVert_2,
\end{equation}
where $sign(\cdot)=1$ if $\bm{d^{+}_{_\parallel q}} - \bm{d_{_\parallel q}}$ and $\bm{d^{+}_{_\parallel q}}$ are in the same direction and $-1$ otherwise, $\lVert\cdot\rVert_2$ is the $\ell_2$ norm, $d$ can be any  document and $\smash{d_q^+}$ is a relevant document for $q$, and $\smash{\bm{d_{_\parallel q}}}$ is defined as
\begin{equation}
\small
\setlength{\abovedisplayskip}{1pt}
\setlength{\belowdisplayskip}{1pt}
\bm{d_{_\parallel q}}=\frac{(\bm{d} \cdot \bm{q})*\bm{q}}{|\bm{q}|*|\bm{q}|}.
\end{equation}
The $PSS$ measures the superiority of $\smash{d_q^+}$ being more relevant to the query $q$ than $d$, by comparing the differences between their information related to $q$. Therefore, a higher $PSS$ value suggests that $d$ is less likely to be an unlabeled relevant sample for $q$.

\noindent \textbf{Definition 2 (Inter Sample Diversity).} 
For a given query $q$, the \textit{inter sample diversity} between $d$ and a document set $D$ is
\begin{equation}
\small
\setlength{\abovedisplayskip}{0pt}
\setlength{\belowdisplayskip}{0pt}
ISD\left(d, D; q\right)=\frac{1}{|D|}\sum_{d^{\prime} \in D}\left\lVert\bm{d_{\perp q}}-\bm{d_{\perp q}^{\prime}}\right\rVert_{2},
\end{equation}
where $\bm{d_{\perp q}}=\bm{d}-\bm{d_{_\parallel q}}.$
The $ISD$ measures the diversity of document $d$ relative to the document set $D$, by comparing the differences between the information unrelated to $q$ among the documents.

Based on the two defined criteria, we introduce each step of the model learning, taking $(q, \smash{d_q^+})\! \sim\! \mathcal{C}_0$ in session $t$ as an example.

\vspace{0.8mm}
\noindent\textbf{STEP 1: New Data Selection.}
For using the new data to adapt to the current session $t$, we have the following principles: 
(1) Documents that are likely to be unlabeled positives should be avoided during selection, since mistakenly identifying relevant documents for model learning could cause serious damage to the performance.  
(2) The selected documents should be the negatives that can support the model to learn the decision boundary (we refer to them as support negatives). Such documents should be not trivial for the model to differentiate.  
(3) The selected documents should have minimum redundancy.  
With these principles, we propose the following selection strategy for the new data.

We first retrieve top results for $q$ from the new-coming document collection $\smash{\mathcal{D}_t}$ with BM25 to filter out massive non-informative samples, and obtain its potential support samples $\smash{D^S_q}$.
Then, based on the defined $PSS$ and $ISD$ criteria, we adaptively select $n_1$ diverse support negatives from $\smash{D^{S}_q}$ with:
\begin{equation}
\small
\label{eq:new_data_select}
\setlength{\abovedisplayskip}{1pt}
\setlength{\belowdisplayskip}{1pt}
D_{\mathit{q}}^{\mathit{new}} \!=\! \bigg\{ \underset{d \in D^{S}_q}{\operatorname{\arg \max}}^{(n_1)}{\ }  \alpha \! \cdot \! PSS(d, d_q^+; q) \! +\! (1\!-\!\alpha) \! \cdot \! ISD(d, D_q^{S}; q)\bigg\},
\end{equation}
where the embedding $\bm{q}$ and $\bm{d}$ used to calculate $PSS$ and $ISD$ are obtained from the latest model $\smash{f_t}$.
The $PSS$ component helps to bypass unlabeled relevant documents and the $ISD$ component prefers the samples that are distinct from the majority.
We use a hyper-parameter $\alpha$ to reconcile the two measures.
With Eq.~(\ref{eq:new_data_select}), we select $n_1$ new documents $\smash{D_{\mathit{q}}^{\mathit{new}}}$ from $\smash{\mathcal{D}_t}$ that satisfy the aforementioned three principles.
These selected samples are reserved for model updating and also added to the temporary memory $\smash{M_c}$ as candidates to update the memory $\smash{M_{t\!-\!1}}$.

\vspace{0.6mm}
\noindent\textbf{STEP 2: Memory Data Selection.}
To prevent the model from forgetting old knowledge when learning from the new data $\smash{\mathcal{D}_t}$, we also select replay samples for model training from $\smash{M_{t\!-\!1}}$ that are: 
(1) pivotal for learning the historical versions of the model, i.e., historical support samples;
(2) not redundant with each other for efficiency concerns;
(3) different from the selected samples in $\smash{D_{\mathit{q}}^{\mathit{new}}}$  to better balance the acquired knowledge and new knowledge.

With the memory updating strategy in STEP 4, the samples in $\smash{M_{t\!-\!1}}$ already satisfy the first two desiderata.
To filter with the third principle, we select $\smash{n_2}$ documents $\smash{D_{\mathit{q}}^{\mathit{mem}}}$ from $\smash{M_{t\!-\!1}}$ that have the maximum $ISD$ score regarding $\smash{D_{\mathit{q}}^{\mathit{new}}}$:
\begin{equation}
\small
\setlength{\abovedisplayskip}{0pt}
\setlength{\belowdisplayskip}{0pt}
D_{\mathit{q}}^{\mathit{mem}}=\bigg \{\underset{d \in D^O_q}{\operatorname{\arg \max}}^{(n_2)}{\ } ISD(d, D_{\mathit{q}}^{\mathit{new}}; q) \bigg \},
\end{equation}
where $\smash{D^O_q}$ denotes the stored old documents for $q$ in the memory buffer $\smash{M_{t\!-\!1}}$.
Note that for computing the $ISD$ score, the embedding $\bm{d}$ of memory samples are the existing ones computed in previous sessions when the learning is backward-compatible. Otherwise, the embedding is obtained using the latest model $\smash{f_t}$.
In the rest of this paper, we adopt the same approach to compute $ISD$, and we will omit this reminder unless there are special circumstances.

\noindent\textbf{STEP 3: Model Update.}
With the selected new documents $\smash{D_{\mathit{q}}^{\mathit{new}}}$ (from STEP 1) and memory documents $\smash{D_{\mathit{q}}^{\mathit{mem}}}$ (from STEP 2), we can update a standard retrieval model from $\smash{f_{t\!-\!1}}$ to $\smash{f_t}$.
Without loss of generality, the retrieval model $\smash{f_t}$ can be formalized as,
\begin{equation}
\small
\setlength{\abovedisplayskip}{0pt}
\setlength{\belowdisplayskip}{0pt}
f_t(q,d) = \langle \operatorname{E}_{t}^{q}(q), \operatorname{E}_{t}^{d}(d) \rangle,
\end{equation}
where $\smash{\operatorname{E}_{t}^{q}}$ and $\smash{\operatorname{E}_{t}^{d}}$ are the query and document encoders, and the dot-product function is used to calculate the relevance score based on their embeddings.
For model training, we use the standard contrastive learning objective~\cite{xiong2020approximate,karpukhin2020dense} to compute the loss for the positive document $\smash{d_q^+}$ (no compatibility is ensured)\footnote{Here, we omit the in-batch negatives in Eq.~(\ref{contrastive_learning}) for brevity.}: 
\begin{equation}
\small
\label{contrastive_learning}
\setlength{\abovedisplayskip}{0pt}
\setlength{\belowdisplayskip}{0pt}
L_t^{\mathit{no-com}} = - \log \frac{\exp (f_t(q,d_q^+))}{\sum_{d \in \{d_q^+\} \cup D_{\mathit{q}}^{\mathit{new}} \cup D_{\mathit{q}}^{\mathit{mem}}} \exp (f_t(q,d))}.
\end{equation}
When the model updating is not backward-compatible for document representations, we need to re-embed all the documents up to the $t$-th session, i.e., $\smash{D_{0:t}}$, with $\smash{f_t}$ to rebuild the retrieval index.
To eliminate the need for re-inferring embeddings of old documents, we can replace the learning objective in Eq.~(\ref{contrastive_learning}) with the backward-compatible  learning objective in Section~\ref{sec:compatible_representation_learning}. 

\noindent\textbf{STEP 4: Memory Update.}
In practice, the memory buffer size is often limited to ensure efficiency in selecting replay samples, even though it does not impose a heavy storage burden. 
Given the limited budget $n$ for each query, selecting which samples to include or replace in the memory is critical. 
We consider two principles to populate the memory: (1) The sample should have a strong impact on the learning of the decision boundary; (2) The redundancy between stored samples should be minimized.
In contrast to most work that updates the memory in each training step~\cite{aljundi2019mir, chaudhry2019er}, we delay the memory update until after the completion of model updating in each session in order not to occupy the limited slots in the buffer.

To preserve important information of the current session $t$ for the future, we follow the first principle and consider only the support samples in $\smash{M_c}$ as candidates to update $\smash{M_{t\!-\!1}}$.
We calculate the $ISD$ score of documents in $\smash{M_{t\!-\!1}}$ and $\smash{M_c}$ regarding $k$ randomly-sampled anchor documents in $\smash{M_{t\!-\!1}}$, and use the new documents with the maximum diversity to replace $\smash{n_3}$ memory samples with the minimum diversity.
Finally, we empty the temporary memory buffer to prepare for the next session.
Note that, for the initial session ($t\!=\!0$), we use reservoir sampling~\cite{chaudhry2019er} to fill the memory. 

\vspace{-3mm}
\subsection{Backward-compatible Learning}  \label{sec:compatible_representation_learning}
To save the cost of repeated embedding computation, it is desirable for the model updating to ensure backward-compatibility of document representations. 
It means that existing embeddings for $\smash{\mathcal{D}_{0:t\!-\!1}}$ do not need to be updated, and only the embeddings of new documents in $\smash{\mathcal{D}_t}$ are computed with the lastest model $\smash{f_t}$ to update the index.
We first introduce a vanilla method that can ensure backward compatibility, and two auxiliary alignment objectives for effective compatible learning.

\vspace{0.2mm}
\noindent\textbf{Vanilla Compatible Learning.}
A straightforward approach is to  optimize a new contrastive learning loss by fixing the embeddings of previous documents (i.e., the positive sample and the memory samples selected in the current training):
\begin{equation}
\small
\label{rank_loss}
\setlength{\abovedisplayskip}{1pt}
\setlength{\belowdisplayskip}{1pt}
L_t^{\mathit{rank}} \!=\! - \!\log \frac{\exp (\langle\operatorname{E}_{t}^{q}(q), \bm{d_q^+}\rangle)}{Z},
\end{equation}
where $Z$ is a normalization term:
\begin{equation}
\small
\setlength{\abovedisplayskip}{1pt}
\setlength{\belowdisplayskip}{1pt}
Z = \sum_{d \in \{d_q^+\} \cup D_{\mathit{q}}^{\mathit{mem}}} \!\exp (\langle\operatorname{E}_{t}^{q}(q), \bm{d}\rangle) + \sum_{d \in D_{\mathit{q}}^{\mathit{new}}} \!\exp (f_t(q,d)).
\end{equation}
The Eq.~(\ref{rank_loss}) optimizes the model on the new data and existing document embeddings in a unified space to ensure compatibility.
However, since all the new samples in $\smash{D_{\mathit{q}}^{\mathit{new}}}$ are negatives and only the embeddings of new samples are learnable, the model could easily learn the wrong correlation between a document being in the new distribution and it being irrelevant, leading to significant performance regression (see the experimental results in Section~\ref{sec:results}).
In order to facilitate effective backward-compatible representation learning, we introduce two auxiliary alignment objectives.

\vspace{0.2mm}
\noindent\textbf{Embedding-aligned Learning.} 
As in~\cite{shen2020towards}, a common approach to ensure backward-compatible model updating is to minimize the $\smash{\ell_{2}}$ distance between the embeddings of previous documents (i.e., $\smash{\{d_q^+\} \!\cup\! D_{\mathit{q}}^{\mathit{mem}}}$) calculated with the new model $\smash{f_t}$ and their existing embeddings:
\begin{equation}
\small
\label{emb_alignment}
\setlength{\abovedisplayskip}{-1pt}
\setlength{\belowdisplayskip}{0pt}
L_t^{\mathit{align_e}}=\sum_{d \in \{d_q^+\} \cup D_{\mathit{q}}^{\mathit{mem}}} \frac{1}{2}\left\lVert\operatorname{E}_{t}^{d}(d) - \bm{d}\right\rVert_{2}^{2}.
\end{equation}
By guiding the model to encode the old documents similarly to their existing embeddings, it could urge the model to learn decent document representations instead of blindly demoting new documents. 
However, this pointwise alignment is too strict for the model to adapt to the new documents sufficiently.

\vspace{0.2mm}
\noindent\textbf{Ranking-aligned Learning.}
To relax the constraint on the model to learn new knowledge, we propose a loose listwise alignment objective. The goal is to minimize the divergence between the predicted distributions of the candidate documents calculated based on the existing and currently learned embeddings, i.e., $\smash{p(D \!\mid\! q)}$ and $\smash{p^{\prime}(D \!\mid\! q)}$, respectively:
\begin{equation}
\small
\label{rank_align}
\setlength{\abovedisplayskip}{1pt}
\setlength{\belowdisplayskip}{2pt}
L_t^{\mathit{align_r}}=\operatorname{KL}\left(p(D \!\mid\! q) \| \ p^{\prime}(D \!\mid\! q)\right)
=\sum_{d \in D} p(d \!\mid\! q) \log \frac{p(d \!\mid\! q)}{p^{\prime}(d \!\mid\! q)},
\end{equation}
where $\smash{D = \{d_q^+\} \cup D_{\mathit{q}}^{\mathit{mem}} \cup D_{\mathit{q}}^{\mathit{new}}}$, and 
\begin{small}
\setlength{\abovedisplayskip}{5pt}
\setlength{\belowdisplayskip}{2pt}
\begin{flalign}
\label{rank_align_old}
\quad &p(d \!\mid\! q) = \left\{\begin{array}{ll}
\!\frac{\exp (f_t(q,d))}{Z} & \text{if } d \! \in \!D_{\mathit{q}}^{\mathit{new}} \\ 
\!\frac{\exp (\langle\operatorname{E}_{t}^{q}(q), \bm{d}\rangle)}{Z} & \text {if } d \!\in\! \{d_q^+\} \cup D_{\mathit{q}}^{\mathit{mem}} \end{array}\right. ,&
\end{flalign}
\end{small}
\begin{small}
\setlength{\abovedisplayskip}{2pt}
\setlength{\belowdisplayskip}{3pt}
\begin{flalign}
\label{rank_align_new}
\quad &p^{\prime}(d \!\mid\! q) = \frac{\exp (f_t(q,d))}{\sum_{d \in D} \exp (f_t(q,d))}.&
\end{flalign}
\end{small}
\hspace{-1mm}The probability distribution $\smash{p(D \!\mid\! q)}$ represents the model inference when backward compatibility is enabled, and $\smash{p^{\prime}(D \!\mid\! q)}$ represents the model predictions without compatible learning where all the embeddings need to be learned.
In contrast to the pointwise embedding alignment, this ranking-based alignment not only allows more flexible exploration in the representation space but also facilitates bidirectional supervision for model learning:  
1) $\smash{p^{\prime}(D \!\mid\! q)}$ can adapt the model better to the new data since the embeddings of candidates are all currently learned including the new ones. So it could guide  $\smash{p(D \!\mid\! q)}$, the backward-compatible inference we finally use, to better acquire new knowledge. 
2) In $\smash{p(D \!\mid\! q)}$, since the positive document and memory negative samples are ranked based on their existing embeddings up until session $t\!-\!1$, $\smash{p(D \!\mid\! q)}$ captures their relative rankings from the model at session $t\!-\!1$. This older model has seen the negatives from session 1 to $t\!-\!1$ including the ones that have been removed from the memory, which could cover various types of negatives. Hence, by aligning with $\smash{p(D \!\mid\! q)}$, $\smash{p^{\prime}(D \!\mid\! q)}$ can learn from the older more experienced model. 
Given the above mutual supervisions between the new and old model, our proposed ranking alignment objective ensures the representation compatibility without compromising the model performance, obtaining even better results (see more analysis in Section~\ref{sec:analysis_compatible}).


\vspace{0.3mm}
\noindent\textbf{Overall Compatible Learning Objective.} The final training objective to enable backward compatibility is the combination of the vanilla ranking loss and the alignment loss:
\begin{equation}
\label{overall}
\setlength{\abovedisplayskip}{1pt}
\setlength{\belowdisplayskip}{1pt}
L_t^{\mathit{com}} = L_t^{\mathit{rank}} + \lambda \cdot  L_t^{\mathit{align}},
\end{equation}
where $\smash{L_t^{\mathit{align}}}$ is either $\smash{L_t^{\mathit{align_e}}}$ or $\smash{L_t^{\mathit{align_r}}}$, and $\lambda$ is a hyper-parameter to control the effect of the alignment regularization.

\vspace{-4mm}
\section{Experimental Settings}

\subsection{Benchmark Construction}
There are no publicly available datasets that could show the continuous growth of documents in realistic retrieval scenarios, potentially with distribution drift, booming events, and newly emerged relevant documents to previous queries.
Thus, we build two benchmarks, i.e., LL-LoTTE and LL-MultiCPR, based on two retrieval datasets LoTTE~\cite{santhanam2021colbertv2} and Multi-CPR~\cite{long2022multi}, to simulate the scenario with the aforementioned properties through the following steps:

\textit{\textbf{Preprocessing.}}
LoTTE and Multi-CPR are two retrieval datasets that consist of 5 and 3 domains with separate subsets of documents and queries respectively. 
For each domain of the two datasets, we merge all the data and re-split them  into train/dev/test sets with a ratio of 0.7:0.15:0.15 for LoTTE and 0.9:0.05:0.05 for Multi-CPR.
Table~\ref{tab:original_dataset} lists the statistics of the two datasets.


\textit{\textbf{Session Partitioning.}}
We build an initial collection $\smash{\mathcal{D}_0}$ and 3 upcoming sessions with different document distributions for both LL-LoTTE and LL-MultiPCR. 
In LL-LoTTE, we use technology and writing as the common domains where documents emerge evenly over time, and lifestyle, recreation, and science as the booming domains in each of the upcoming sessions. 
We keep 70\% and 40\% of the random documents from the common and booming domains respectively in $\smash{\mathcal{D}_0}$.
Next, we construct 3 corpora $\smash{\{\mathcal{D}_1, \mathcal{D}_2, \mathcal{D}_3\}}$ for the following three sessions. Each corpus consists of 10\%, 50\%, and 5\% of documents from the two common domains, a booming domain, and the remaining two domains respectively. 
With the documents in each session, we collect their connected queries from the new split train/dev/test sets of LoTTE to construct the training dataset and dev/test sets.
Note that, under the setting where new-coming documents have no labels, the labeled relevant query-document pairs for model training remains the $\smash{\mathcal{C}_0}$, but the dev/test sets $\smash{\mathcal{Q}_{t}^{\text{dev}}}$ and $\smash{\mathcal{Q}_{t}^{\text{test}} (\forall t \!\in\!\{0, \cdots, 3\})}$ can have more annotated relevant documents for  evaluation.
In LL-MultiCPR, similar to LL-LoTTE, we choose e-commerce as the common domain,  medical and entertainment as the booming domains for Session 1 and 2 respectively. Since there are only three domains, Session 3 has no booming domains and simply includes the remaining documents.

\textit{\textbf{Postprocessing.}} 
In LoTTE, almost all relevant documents of each query have positive labels. This makes it hard to simulate the realistic scenario where quite a few relevant documents to training queries may appear in the upcoming sessions and not be labeled. To overcome this issue, we collect extra pseudo-relevant documents for training queries using OpenAI API (text-davinci-003), and distribute these unlabeled documents to each coming session with the same sampling ratios in session partitioning. 
Specifically, we use two types of instructions for pseudo-relevant document generation: (1) ``Given a question \{$q$\} and a relevant document \{$d_q^+$\}, please generate 5 other relevant documents."; (2) ``Given a document \{$d_q^+$\}, please rephrase it.".
Through this process, we obtain approximately 18.5 documents for each training query\footnote{We perform quantitative analysis on these generated pseudo documents to ensure that they are of high quality. It shows that 63\% of them can be retrieved in the top-200 results of BM25 for training queries in each upcoming session.}.
For LL-MultiCPR, we have not conducted post-processing since there are sizable unlabeled relevant documents in Multi-CPR (see~\cite{long2022multi}). 

Table~\ref{tab:ll-lotte} lists the statistics of the final LL-LoTTE and LL-MultiCPR datasets. 
Following similar steps, other existing retrieval datasets can also be transformed to evaluate lifelong learning of first-stage retrieval. When there are no explicitly separate domains, topic clustering could be applied for simulation and we leave such investigation for future research.

\renewcommand{\arraystretch}{0.8}
\begin{table}[!t]
\large
\setlength\tabcolsep{2.4pt}
\setlength{\abovecaptionskip}{-1pt}
\setlength{\belowcaptionskip}{0pt}
  \caption{Statistical information of LoTTE and Multi-CPR.}
  \label{tab:original_dataset}
  \begin{tabular}{lrrrrr}
    \toprule
    domain & \#query & \#document & len\_q & len\_d & \#qrels \\
    \midrule
    \textbf{\textit{\small LoTTE}} & & & & \\
    \texttt{technology} &5519 &1,914,731 &9.00 &124.69 &6.59  \\ 
    \texttt{writing} &5571 &477,066 &9.11 &171.39 &5.89  \\ 
    \texttt{lifestyle} &5156 &388,354 &10.04 &166.65 &5.10  \\ 
    \texttt{recreation} &5491 &430,000 &9.16 &193.16 &4.27  \\ 
    \texttt{science} &5185 &2,037,806 &9.08 &139.83 &5.98  \\ 
    \midrule
    \textbf{\textit{\small Multi-CPR}} & & & & \\
    \texttt{e-commerce} &101,000 &1,002,822 &6.90 &32.96 &1.0  \\ 
    \texttt{medical} &100,999 &959,526 &17.07 &121.90 &1.0  \\ 
    \texttt{entertainment} &101,000 &1,000,000 &7.41 &27.45 &1.0  \\
    \bottomrule
  \end{tabular}
\vspace{-4mm}
\end{table}

\renewcommand{\arraystretch}{0.8}
\begin{table}[!t]
\large
\setlength\tabcolsep{5pt}
\setlength{\abovecaptionskip}{-1pt}
\setlength{\belowcaptionskip}{-2pt}
  \caption{Statistics of LL-LoTTE and LL-MultiCPR datasets.}
  \label{tab:ll-lotte}
  \begin{tabular}{lrrrr}
    \toprule
     & Session$_0$ & Session$_1$ & Session$_2$ & Session$_3$ \\
    \midrule
    \textbf{\textit{\small LL-LoTTE}} & & & & \\
    \#document &2,816,720  &654,266  &670,026  &1,405,225   \\ 
    \#train\_q &16,147  &-  &- &-  \\ 
    \#dev\_q &3449  &1681  &1750  &1666  \\ 
    \#test\_q &3448  &1707  &1752  &1700  \\ 
    \#test\_qrels &4.16 &6.79  &7.55 &8.31  \\
    \midrule
    \textbf{\textit{\small LL-MultiCPR}} & & & & \\
    \#document &1,486,184  &630,545  &648,307  &198,310   \\ 
    \#train\_q &136,282  &-  &- &-  \\ 
    \#dev\_q &7551  &3340  &3270  &989  \\ 
    \#test\_q &7653  &3242  &3223  &1032  \\ 
    \#test\_qrels &1.0 &1.0  &1.0 &1.0  \\
    \bottomrule
  \end{tabular}
\vspace{-5mm}
\end{table}

\vspace{-2mm}
\subsection{Evaluation Metrics}\label{sec:metrics}
We define metrics to evaluate lifelong learning methods for first-stage retrieval.
Considering the realistic scenario, for each session, we care more about the retrieval performance on the queries in the current session.
Let $\smash{p_{i,j}}$ be the retrieval performance evaluated on the test queries of session $j$ (i.e, $\smash{\mathcal{Q}_{j}^{\text{test}}}$) over the document collection $\smash{\mathcal{D}_{0:j}}$ after the learning of session $i$, and $\smash{p_{i,j}}$ can be measured using any common retrieval metric like Recall or MRR.
We take the performance at session $t$, namely $\smash{P_t}$, and average performance over all coming sessions, namely $AP$, to compare various methods:
\begin{equation}
\small
\setlength{\abovedisplayskip}{-1pt}
\setlength{\belowdisplayskip}{-1pt}
\mathrm{P_t} = p_{t, t},  \qquad
\mathrm{AP} = \frac{1}{T} \sum_{t=1}^{T} p_{t, t}.
\end{equation}
Following~\cite{mai2022online}, we also apply auxiliary metrics to assess how fast a model learns (Training Time), how much the model forgets ($\mathrm{Forget_t}$), and how well the model transfers knowledge from one session to future sessions ($\mathrm{FWT}$).
Formally, they are defined as:
\begin{equation}
\small
\setlength{\abovedisplayskip}{0pt}
\setlength{\belowdisplayskip}{0pt}
\mathrm{Forget_t} \!=\!\frac{1}{t} \sum_{j=0}^{t-1} \max _{l \in\{0, \ldots, t-1\}}\left(p_{l, j}\!-\!p_{t, j}\right),  \quad
\mathrm{FWT} \!=\!\frac{\sum_{i=1}^{j-1} \sum_{j=2}^{T} p_{i, j}}{\frac{T(T-1)}{2}}.
\end{equation}

To instantiate the above metrics in our work, we consider the evaluation method of the original LoTTE~\cite{santhanam2021colbertv2} and Multi-CPR~\cite{long2022multi}. 
Besides Recall (R@$N$), Success (S@$N$) and Mean Reciprocal Rank (MRR@$N$) are used in LoTTE and Multi-CPR respectively. Following the official cutoffs for $N$, we show the lifelong learning performance on the above defined metrics regarding S@5 and R@100 for LL-LoTTE, and MRR@10 and R@1000 for LL-MultiCPR.

\vspace{-3mm}
\subsection{Baselines}\label{sec:baselines}
We consider two types of baselines for comparison: 

\noindent\textbf{Memory-based Methods:} 
(1) ER~\cite{chaudhry2019er} applies random sampling for memory data selection and reservoir sampling for memory update. Despite its simplicity, ER outperforms many complex lifelong learning methods~\cite{mai2022online}.
(2) MIR~\cite{aljundi2019mir} chooses replay samples according to their loss increment regarding the updated model learned on the new data, and also uses reservoir sampling for memory update.
(3) GSS~\cite{aljundi2019gss} has the same memory data selection strategy as ER but refines the memory update by trying to diversify the samples in the memory buffer based on their gradients. However, it incurs huge computation costs.
(4) OCS~\cite{yoon2021ocs} is one of the latest methods for lifelong learning  containing noisy data. It selects high-affinity samples to previous data based on their gradients for model and memory update.

\noindent\textbf{Naive Methods:}
(1) \textit{Initial} conducts no model updating and uses the model trained in the initial session for the retrieval in the upcoming sessions.
(2) \textit{Incre-train} initializes the model training from the previous session and updates it with the new data in the current session. 
(3) \textit{Retrain} trains the model from scratch in each session using the whole available data until that session.

To see the separate effect of our proposed data selection strategy and ranking alignment objective, we compare our method with the baselines both without and with backward-compatible representation learning (based on Eq.~(\ref{contrastive_learning}) or Eq.~(\ref{overall})).
Note that: 
(1) \textit{Initial} has the same performance under the two settings since the model is not updated; 
(2) \textit{Retrain} works only without backward compatibility since the model is retrained from scratch in each session.
For the comparisons with the backward-compatibility constraint, we equip the baselines with the embedding alignment objective in Eq.~(\ref{overall}), and our model \modelname{} uses no or one of the two alignment objectives, named as \textbf{\modelname{}$_{\mathit{vanilla}}$}, \textbf{\modelname{}$_{\mathit{emb}}$}, and \textbf{\modelname{}$_{\mathit{rank}}$}, respectively.

\renewcommand{\arraystretch}{1.05}
\begin{table*}[!t]
\small
\setlength\tabcolsep{1.2pt}
\setlength{\abovecaptionskip}{1pt}
\centering
\fontsize{9}{9}\selectfont
\caption{Evaluation results on LL-LoTTE and LL-MultiCPR without representation compatibility. 
    Bold and \underline{underline} indicate the best overall and baseline performance. 
    $\ast$ indicates statistically significant improvements over all baselines (p < 0.05).}
\label{tab:performance_comparison_no_compatible}
\begin{threeparttable}
\begin{tabular}{l|cllll cllll|cllll cllll}
\toprule
\multirow{3}*{Method}   &\multicolumn{10}{c|}{LL-LoTTE} &\multicolumn{10}{c}{LL-MultiCPR}  \\
\cmidrule(lr){2-11} \cmidrule(lr){12-21}
 \multicolumn{1}{l|}{}  &\multicolumn{5}{c}{S@5} &\multicolumn{5}{c|}{R@100}  &\multicolumn{5}{c}{MRR@10} &\multicolumn{5}{c}{R@1000}  \\
 \cmidrule(lr){2-6} \cmidrule(lr){7-11} \cmidrule(lr){12-16} \cmidrule(lr){17-21}
 \multicolumn{1}{c|}{}  &\multicolumn{1}{c}{P$_0$} &\multicolumn{1}{c}{P$_1$} &\multicolumn{1}{c}{P$_2$} &\multicolumn{1}{c}{P$_3$} &\multicolumn{1}{c}{AP} &\multicolumn{1}{c}{P$_0$} &\multicolumn{1}{c}{P$_1$} &\multicolumn{1}{c}{P$_2$} &\multicolumn{1}{c}{P$_3$} &\multicolumn{1}{c|}{AP}  &\multicolumn{1}{c}{P$_0$} &\multicolumn{1}{c}{P$_1$} &\multicolumn{1}{c}{P$_2$} &\multicolumn{1}{c}{P$_3$} &\multicolumn{1}{c}{AP} &\multicolumn{1}{c}{P$_0$} &\multicolumn{1}{c}{P$_1$} &\multicolumn{1}{c}{P$_2$} &\multicolumn{1}{c}{P$_3$} &\multicolumn{1}{c}{AP}  \\
\midrule
BM25    &40.0 &45.3 &43.6 &\underline{44.5} &44.5    &47.1 &43.0 &40.2 &\underline{37.6} &40.3  &19.69&14.78&17.04&15.97&15.93  &72.43&51.79&73.35&68.99&64.71 \\
\midrule
Initial    &41.0 &47.4 &44.3 &41.6 &44.4    &48.4 &43.7 &41.7 &35.4&40.3  &25.16&16.22&20.79&19.88&18.96  &83.88&66.66&79.77&78.20&74.88 \\
Incre-train  &/&47.3&45.5&42.1&45.0  &/&43.7&41.9&35.1&40.2 &/&15.32&20.67&19.85&18.61  &/&65.21&78.68&78.07&73.99 \\
Retrain   &/&47.4&44.5&41.0&44.3   &/&43.6&40.2&33.9&39.2  &/&15.51&20.25&19.50&18.42  &/&64.37&78.93&76.65&73.32 \\ 
\midrule
ER   &/&47.8&45.4&42.6&45.3  &/&44.0&42.0&35.3&40.4    &/&16.15&20.87&20.14&19.05  &/&66.93&79.74&78.59&\underline{75.09}  \\  
MIR    &/&\underline{48.7}&\underline{46.1}&43.4&\underline{46.1}  &/&44.2&\underline{42.7}&36.0&\underline{41.0}   &/&16.07&\underline{21.01}&\underline{20.32}&\underline{19.13}  &/&66.90&\underline{79.77}&78.49&75.05  \\ 
GSS   &/&48.3&45.8&43.3&45.8 &/&44.2&42.3&35.4&40.6 &/&\underline{16.43}&20.78&19.95&19.05  &/&\underline{67.02}&79.37&\underline{78.71}&75.03   \\ 
OCS  &/&48.6&\underline{46.1}&43.4&46.0  &/&\underline{44.3}&42.5&35.9&40.9 &/&16.39&20.57&20.22&19.06  &/&66.75&79.46&78.29&74.83      \\ 
\smash{\modelname{}}   &/ &$\bm{50.0}^{\ast}$ &$\bm{48.0}^{\ast}$ &$\bm{46.5}^{\ast}$ &$\bm{48.2}^{\ast}$  &/ &$\bm{45.9}^{\ast}$ &$\bm{44.5}^{\ast}$ &$\bm{38.2}^{\ast}$ &$\bm{42.9}^{\ast}$     &/ &$\bm{17.25}^{\ast}$ &$\bm{22.34}^{\ast}$ &$\bm{21.57}^{\ast}$ &$\bm{20.39}^{\ast}$  &/ &$\bm{68.69}^{\ast}$ &$\bm{80.55}^{\ast}$ &$\bm{80.17}^{\ast}$ &$\bm{76.47}^{\ast}$    \\ 
\bottomrule
\end{tabular}
\end{threeparttable}
\vspace{-3mm}
\end{table*}

\renewcommand{\arraystretch}{1.05}
\begin{table*}[!t]
\small
\setlength\tabcolsep{3.3pt}
\setlength{\abovecaptionskip}{1pt}
\centering
\fontsize{9}{9}\selectfont
\caption{Evaluation results on LL-LoTTE and LL-MultiCPR with representation compatibility. 
    Bold and \underline{underline} indicate the best overall and baseline performance. 
    $\ast$ indicates statistically significant improvements over all baselines (p < 0.05).}
\label{tab:performance_comparison_with_compatible}
\begin{threeparttable}
\begin{tabular}{l|llll llll|llll llll}
\toprule
\multirow{3}*{Method}   &\multicolumn{8}{c|}{LL-LoTTE} &\multicolumn{8}{c}{LL-MultiCPR}  \\
\cmidrule(lr){2-9} \cmidrule(lr){10-17}
 \multicolumn{1}{c|}{}  &\multicolumn{4}{c}{S@5} &\multicolumn{4}{c|}{R@100}  &\multicolumn{4}{c}{MRR@10} &\multicolumn{4}{c}{R@1000}  \\
 \cmidrule(lr){2-5} \cmidrule(lr){6-9} \cmidrule(lr){10-13} \cmidrule(lr){14-17}
 \multicolumn{1}{c|}{}  &\multicolumn{1}{c}{P$_1$} &\multicolumn{1}{c}{P$_2$} &\multicolumn{1}{c}{P$_3$} &\multicolumn{1}{c}{AP} &\multicolumn{1}{c}{P$_1$} &\multicolumn{1}{c}{P$_2$} &\multicolumn{1}{c}{P$_3$} &\multicolumn{1}{c|}{AP} &\multicolumn{1}{c}{P$_1$} &\multicolumn{1}{c}{P$_2$} &\multicolumn{1}{c}{P$_3$} &\multicolumn{1}{c}{AP} &\multicolumn{1}{c}{P$_1$} &\multicolumn{1}{c}{P$_2$} &\multicolumn{1}{c}{P$_3$} &\multicolumn{1}{c}{AP}  \\
\midrule
Initial    &\underline{47.4} &\underline{44.3} &\underline{41.6} &\underline{44.4}    &\underline{43.7} &\underline{41.7} &\underline{35.4} &\underline{40.3}  &\underline{16.22} &\underline{20.79} &\underline{19.88} &\underline{18.96}  &\underline{66.66} &\underline{79.77} &\underline{78.20} &\underline{74.88} \\
Incre-train  &44.9&42.2&39.0&42.0   &41.1&39.1&33.9&38.0 &8.85&14.54&12.11&11.83  &57.77&76.02&72.45&68.75 \\
\midrule
ER   &45.2&41.6&37.8&41.5  &41.4&38.3&32.8&37.5    &9.31&13.61&11.66&11.53  &57.80&75.18&71.03&68.00   \\ 
MIR   &45.2&42.2&38.3&41.9  &41.5&38.5&33.0&37.7   &9.23&13.62&11.50&11.45  &57.87&75.09&70.93&67.96  \\ 
GSS   &45.1&42.1&38.2&41.8  &41.4&38.8&33.1&37.8    &9.29&13.46&11.32&11.36  &57.95&75.37&71.14&68.15  \\
OCS  &45.1&42.3&38.3&41.9  &41.5&38.8&33.2&37.8  &9.28&13.16&11.42&11.29  &57.77&75.01&70.89&67.89  \\ 
\smash{\modelname{}$_{\textit{vanilla}}$}  &40.3&36.8&33.3&36.8  &31.0&27.0&22.0&26.7  &3.31&9.64&7.75&6.90  &23.87&47.16&41.67&37.57  \\ 
\smash{\modelname{}$_{\textit{emb}}$} &46.3&43.8&38.8&43.0  &42.4&38.7&33.4&38.2   &9.38&14.20&12.27&11.95  &58.11&75.80&72.38&68.76  \\  
\smash{\modelname{}$_{\textit{rank}}$} &$\bm{50.6}^{\ast}$ &$\bm{47.3}^{\ast}$ &$\bm{44.6}^{\ast}$ &$\bm{47.5}^{\ast}$  &$\bm{46.9}^{\ast}$ &$\bm{44.1}^{\ast}$ &$\bm{37.8}^{\ast}$ &$\bm{42.9}^{\ast}$   &$\bm{22.61}^{\ast}$ &$\bm{25.80}^{\ast}$ &$\bm{29.11}^{\ast}$ &$\bm{25.84}^{\ast}$  &$\bm{70.64}^{\ast}$ &$\bm{80.05}^{\ast}$ &$\bm{80.91}^{\ast}$ &$\bm{77.20}^{\ast}$ \\ 
\bottomrule
\end{tabular}
\end{threeparttable}
\vspace{-3mm}
\end{table*}

\renewcommand{\arraystretch}{1.15}
\setlength\tabcolsep{1.3pt}
\begin{table}[!t]
\small
\setlength{\abovecaptionskip}{1pt}
  \centering
  \fontsize{9}{9}\selectfont
  \caption{Evaluation results of the last session ($P_3$) with different buffer size on LL-LoTTE and LL-MultiCPR. All the methods run with representation compatibility.}
  \label{tab:buffer_size}
  \begin{threeparttable}
  \begin{tabular}{l|cccc|cccc}
    \toprule
     \multirow{3}{*}{Method}  
     &\multicolumn{4}{c|}{LL-LoTTE}
      &\multicolumn{4}{c}{LL-MultiCPR}   \\
     \cmidrule(lr){2-5} \cmidrule(lr){6-9}
     &\multicolumn{2}{c}{$n$=30} &\multicolumn{2}{c|}{$n$=100} &\multicolumn{2}{c}{$n$=10} &\multicolumn{2}{c}{$n$=30}  \\
     \cmidrule(lr){2-3} \cmidrule(lr){4-5} \cmidrule(lr){6-7} \cmidrule(lr){8-9}
     &\small S@5 &\small R@100 &\small S@5 &\small R@100 &\small MRR@10 &\small R@1000 &\small MRR@10 &\small R@1000 \\
    \midrule
    ER &37.8 &32.8 &38.1 &32.9 &11.66 &71.03 &11.71 &71.32 \cr
    MIR &38.3 &33.0 &37.9 &33.0 &11.50 &70.93 &11.72 &71.41 \cr
    GSS &38.2 &33.1 &38.4 &33.1 &11.32 &71.14 &11.44 &71.33 \cr
    OCS &38.3 &33.2 &38.8 &33.1 &11.42 &70.89 &11.43 &71.01 \cr
    \smash{\modelname{}$_{\textit{rank}}$} &44.6 &37.8 &45.2 &38.2 &29.11 &80.91 &30.27 &81.30 \cr
    \bottomrule
  \end{tabular}
\end{threeparttable}
\vspace{-7mm}
\end{table}

\vspace{-2mm}
\subsection{Implementation Details} 
We implement the retrieval model with DPR~\cite{karpukhin2020dense}, and the parameters are initialized with BERT-base released by Google. 
The hyper-parameters in baselines and our method are tuned on the dev set.
For LL-LoTTE, we truncate the input query and passage to a maximum of 32 and 256 tokens respectively. We train retrieval models with BM25 top-500 results for the initial session and top-200 results for the upcoming sessions, and the key hyper-parameters of BM25 are tuned to $k_1 \!=\! 0.80$ and $b \!=\! 0.72$. We use a batch size of 96, and a learning rate of 5e-6 and 1e-6 for the initial session and upcoming sessions respectively. 
For LL-MultiCPR, we set the query and passage length to 32 and 128 respectively. We train the initial session and upcoming sessions with BM25 top-500 results, with $k_1 \!=\! 0.20$ and $b \!=\! 0.72$. We use a learning rate of 1e-5 and 3e-6 for the initial and upcoming sessions respectively, and a batch size of 192.
For the two datasets, we pair each positive document with 5 negatives for training, including 3 new documents and 2 memory documents. 
For data selection, we upsample a subset with twice the desired number of documents in each training step, instead of the entire collection, to save the computation cost.
For memory update, we set the number of anchor documents and replaced documents are 1/3 of the memory buffer size $n$.
We set $\alpha$ to 0.6 and 0.8 for LL-LoTTE and LL-MultiCPR respectively, and $\lambda$ to 1.0 and 3.0 for both.
For each dataset, we set the buffer size $n$ of each training query with two settings that can hold: (1) half of training negatives in the initial session (i.e., 30 for LL-LoTTE and 10 for LL-MultiCPR); (2) total training negatives in all the sessions (i.e., 100 for LL-LoTTE and 30 for LL-MultiCPR). We use the former as the default setting.

We adopt the Transformers for implementations and all experiments run on Nvidia Tesla V100-32GB GPUs.  
Statistically significant differences are measured by a two-tailed t-test.
The datasets and code are available at https://github.com/caiyinqiong/L-2R.

\vspace{-2mm}
\section{Results and Discussion} \label{sec:results}
In this section, we present the experimental results and conduct thorough analysis of \modelname{} to clarify its advantages.

\vspace{-2mm}
\subsection{Main Evaluation}
We compare \modelname{} with all the baselines in Section~\ref{sec:baselines}, and record their results under both settings in Table~\ref{tab:performance_comparison_no_compatible} \& ~\ref{tab:performance_comparison_with_compatible}.

\textbf{Performance without Representation Compatibility.}  
From Table~\ref{tab:performance_comparison_no_compatible}, we find that:
(1) Without special measures for lifelong learning, neural retriever DPR (i.e., \textit{Initial}) shows poorer generalization ability than the term-based retrievers (i.e., BM25), especially when the distribution changes violently. 
For example, DPR outperforms BM25 on LL-LoTTE in Session 0-2 but underperforms  it when massive science documents influx in Session 3 (there are significantly more documents in the science domain than others). 
This observation is consistent with the conclusion in~\cite{ren2022thorough} that neural retrievers are less robust than BM25.
(2) For the methods that learn from new data (i.e., methods except  \textit{Initial}), \textit{Incre-train} performs poorly than memory-based methods, probably because it does nothing to address the catastrophic forgetting issue. 
Additionally, \textit{Incre-train} is not always superior to \textit{Initial}, particularly on LL-MultiCPR.
Apart from the forgetting issue, we believe a potential reason is that the sizable unlabeled relevant documents in the new data could hurt model updating.
(3) It is worth noting that \textit{Retrain} exhibits worse performance, particularly on recall, which deviates from findings in other lifelong learning  tasks like image classification~\cite{aljundi2019mir}. It is probably because the retrained retriever has seen fewer varieties of negative samples and has a higher probability of using emerged unlabeled positive documents for training.
(4) Among the memory-based methods, MIR has the overall best performance on both datasets and OCS can not exceed it. This shows that the gradient-based method to filter out noisy data in OCS does not work effectively for unlabeled relevant documents in the retrieval task.
(5) On both benchmarks, \modelname{} consistently outperforms the baselines in all upcoming sessions. Especially in Session 3 of LL-LoTTE that has violent distribution drift, \modelname{} beats others by a large margin and surpasses BM25. These performance gains confirm the advantages of our proposed data selection strategy in \modelname{}.

\textbf{Performance with Representation Compatibility.} 
The performance of all the methods with representation compatibility is presented in Table~\ref{tab:performance_comparison_with_compatible}. 
Compared to Table~\ref{tab:performance_comparison_no_compatible}, we have the following observations.
From the perspective of \textbf{effectiveness}: 
(1) Adding the embedding alignment to ensure representation compatibility leads to significantly lower model performance, even worse than \textit{Initial}.
It shows that enforced embedding alignment could hurt model learning on new data.
Among these methods, \textit{Incre-train} is hurt the least, probably because the regularization is applied on fewer documents.
(2) For the three variants of \modelname{} with representation compatibility, \modelname{}$_{\textit{vanilla}}$ suffers from model collapse by only optimized with the contrastive learning loss on existing embeddings of previous documents.
By injecting an alignment regularization, the model could be updated more effectively with backward-compatible representations.
(3) It is exciting that \modelname{}$_{\mathit{rank}}$ can significantly exceed \modelname{}$_{\mathit{emb}}$, and even outperform \modelname{} that without representation compatibility in some sessions (e.g., Session 1 of LL-LoTTE and almost all the sessions of LL-MultiCPR). 
It shows that the alignment on predicted ranking lists allows for more flexible  encoder updates than direct embedding alignment, and the prediction results based on existing embeddings (computed by old models) provide beneficial information based on the previously acquired knowledge to guide the model to learn new data (see further analysis in Section~\ref{sec:analysis_compatible}).
From the perspective of \textbf{efficiency}: (1) With representation backward-compatibility, it can save 79\% (2.73M \textit{vs.} 13.16M) and 81\% (1.47M \textit{vs.} 7.85M) of computation costs for inferring document representations than that without compatibility on LL-LoTTE and LL-MultiCPR respectively (accumulated on 3 upcoming sessions).
Overall, these results demonstrate that the ranking alignment objective in \modelname{} could promote both the effectiveness and efficiency of model lifelong learning.

\textbf{Performance with Larger Memory Buffer Size.} 
To investigate the impact of memory buffer size on model performance, we conduct experiments using a larger buffer that can hold all the training samples used in the four sessions. 
Only the results of the last session using different $n$ are reported in Table~\ref{tab:buffer_size} for a clear comparison.
We observe that with a larger buffer size, the performance of \modelname{}$_{\mathit{rank}}$ is further improved, particularly on precision metrics such as S@5 and MRR@10 (e.g., the improvement is 1.3\% on S@5 for LL-LoTTE and 4.0\% on MRR@10 for LL-MultiCPR.).
However, the baseline methods do not benefit as much from a larger memory, probably because ER and MIR store random-sampled documents, and more importantly, all of them cannot filter out the unlabeled positives. 
In contrast, \modelname{} stores diverse support negative samples, thereby making more efficient use of the memory buffer slots.

\vspace{-2mm}
\subsection{Studies on Data Selection Strategy}
We run ablation studies on the data selection strategy   to investigate its impact on model learning. 

For data selection, we define $PSS$ and $ISD$ to measure the likelihood of a document being negative and its diversity relative to others.
We compare \modelname{}$_{\textit{rank}}$ with several ablation variants to verify the effectiveness of our criteria in Table~\ref{tab:ablation_data_selection}:
(1) For the $NewDataSelection$ module, we observe that without the $PSS$ component to filter out unlabeled relevant documents in the new data, the retrieval performance on the two datasets significantly decreases.
Removing $ISD$ also causes a performance drop, especially on recall, since the retriever has seen fewer varieties of negative samples if redundancy among the selected samples is not considered.
These results demonstrate that both criteria are important in selecting new data for the model to adapt to new distributions.
(2) For the $MemoryDataSelection$ module, we remove the $ISD$ component and randomly select replay samples from the memory.
The performance decreases on both datasets,  showing that selecting samples different from the new data for replaying is critical for effective model updating. It is probably because  the cooccurrence of discrepant or even conflicting data encourages the model to deliberate the balance between learning new knowledge and preserving old knowledge.
(3) For the $MemoryUpdate$ module, we remove the $ISD$ component and replace the samples in the memory randomly.
The results show that LL-LoTTE has less regression in performance compared to LL-MultiCPR, probably because LL-MultiCPR uses a smaller buffer size ($n$=10), and storing non-redundant samples becomes more important for it to address the forgetting issue.

\renewcommand{\arraystretch}{0.9}
\setlength\tabcolsep{2pt}
\begin{table}[!t]
\large
\setlength{\abovecaptionskip}{0pt}
  \centering
  \fontsize{9}{9}\selectfont
  \caption{Ablation studies on the data selection strategy in \smash{\modelname{}$_{\textit{rank}}$}. Evaluation results of the last session (\smash{$P_3$}) in LL-LoTTE and LL-MultiCPR are reported.}
  \label{tab:ablation_data_selection}
  \begin{threeparttable}
  \begin{tabular}{ll cc cc}
    \toprule
     \multirow{2}{*}{Module} &\multirow{2}{*}{Strategy}
     &\multicolumn{2}{c}{LL-LoTTE} &\multicolumn{2}{c}{LL-MultiCPR}   \\
     \cmidrule(lr){3-4} \cmidrule(lr){5-6}
     & &S@5 &R@100 &MRR@10 &R@1000  \\
    \midrule
    \multirow{1}*{\smash{\textit{\modelname{}$_{\textit{rank}}$}}} & &44.6 &37.8  &29.11 &80.91  \cr
    \midrule
    \multirow{3}*{\textit{NewDataSelection}}  &-\small$PSS$ &43.8 &37.3  &28.99 &79.65  \cr
    &-\small$ISD$ &44.3 &37.4 &29.14 &80.62  \cr
    &-\text{Both} &43.6 &37.2  &28.82 &79.36  \cr
    \midrule
    \multirow{1}*{\textit{MemoryDataSelection}} &-\small$ISD$ &44.4 &37.1 &28.19 &79.75 \cr
    \midrule
    \multirow{1}*{\textit{MemoryUpdate}} &-\small$ISD$ &44.3 &37.5  &28.59 &80.10 \cr 
    \bottomrule
  \end{tabular}
\end{threeparttable}
\vspace{-4mm}
\end{table}

\renewcommand{\arraystretch}{0.9}
\setlength\tabcolsep{2.2pt}
\begin{table}[!t]
\large
\setlength{\abovecaptionskip}{0pt}
  \centering
  \fontsize{9}{9}\selectfont
  \caption{Investigations on the alignment objectives. Evaluation of each session (\smash{$P_1-P_3$}) in LL-LoTTE are reported.}
  \label{tab:result_old_new}
  \begin{threeparttable}
  \begin{tabular}{ll cc cc cc}
    \toprule
     \multirow{2}{*}{Method} &\multirow{2}{*}{Query}
     &\multicolumn{2}{c}{\modelname{}} &\multicolumn{2}{c}{\modelname{}$_{\mathit{emb}}$}   &\multicolumn{2}{c}{\modelname{}$_{\mathit{rank}}$} \\
     \cmidrule(lr){3-4} \cmidrule(lr){5-6}  \cmidrule(lr){7-8}
     & &S@5 &R@100 &S@5 &R@100 &S@5 &R@100  \\
    \midrule
    \multirow{2}*{Session$_1$}  &\#seen: 1469 &51.1 &43.3 &48.5 &41.5  &50.9 &44.2 \cr
    &\#unseen: 238 &40.3 &57.1 &32.4 &47.5 &48.7 &63.3 \cr
    \midrule
    \multirow{2}*{Session$_2$}  &\#seen: 1525 &48.9 &41.3 &46.6 &37.4  &47.8 &41.6 \cr
    &\#unseen: 227 &39.2 &58.0 &25.1 &47.3  &44.1 &60.6 \cr
    \midrule
    \multirow{2}*{Session$_3$}  &\#seen: 1573 &47.7 &37.5 &40.6 &33.0  &45.3 &36.9 \cr
    &\#unseen: 127 &30.7 &47.6 &17.3 &38.1  &36.2 &50.2 \cr
    \bottomrule
  \end{tabular}
\end{threeparttable}
\vspace{-4mm}
\end{table}

\vspace{-3mm}
\subsection{Studies on Alignment Objectives}\label{sec:analysis_compatible}
We conduct studies on the embedding and ranking alignment objectives to probe their impact on model updating.

\textbf{Performance on Seen and Unseen Queries.}
To understand how the alignment objectives affect model updating, we split the test set of each coming session in LL-LoTTE to previously seen queries and newly unseen queries, and evaluate the performance of \smash{\modelname{}}, \smash{\modelname{}$_{\mathit{emb}}$}, and \smash{\modelname{}$_{\mathit{rank}}$}.
From Table~\ref{tab:result_old_new}, we find that:
(1) The seen queries generally achieve higher S@5 but lower R@100. It is because the seen queries usually have more relevant documents than the unseen queries, which is less favourable for them on recall.
(2) In \smash{\modelname{}$_{\mathit{emb}}$}, both seen and unseen queries experience a significant performance drop compared to \modelname{} that without compatibility. Especially, the drop on unseen queries is more dramatic than that on seen queries. It shows that direct embedding alignment constrains the model to learn new knowledge.
(3) It is interesting that \smash{\modelname{}$_{\mathit{rank}}$} demonstrates improved performance on unseen queries compared to \smash{\modelname{}}. It shows that the ranking results predicted on the old embeddings provide beneficial supervision to the model to learn relevance matching on new data. Moreover, the ranking alignment does not harm seen queries, unless the distribution changes drastically and the model compromises to fit new data (i.e., Session 3).

\begin{figure}[!t]
\setlength{\belowcaptionskip}{-0.4cm}
\setlength{\abovecaptionskip}{-0.1cm}
\includegraphics[scale=0.5]{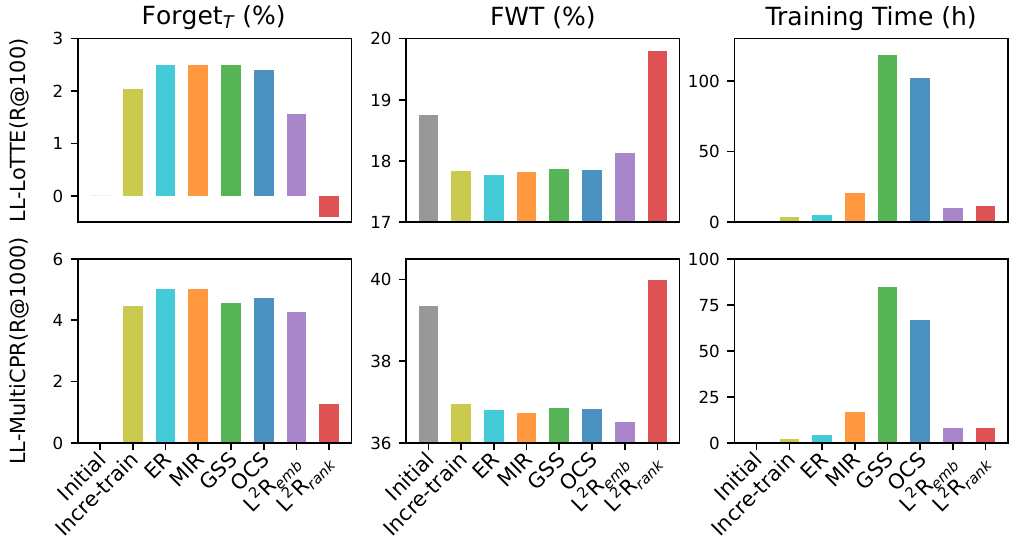}
\caption{Evaluation on auxiliary metrics. Each column denotes a metric and each row denotes a dataset.}
\label{fig:all_metrics}
\end{figure}

\textbf{Performance on Auxiliary Metrics.}
We compare all the methods with representation compatibility on auxiliary metrics, including $\mathrm{Forget_T}$, $\mathrm{FWT}$, and Training Time, to gain insights into the model updating process.
From Figure~\ref{fig:all_metrics}, we observe:
(1) Among all the methods, \modelname{}$_{\mathit{rank}}$ performs best in addressing the catastrophic forgetting issue. Particularly on LL-LoTTE, it has negative values on  $\mathrm{Forget_T}$. 
Apart from the superior memory mechanism in \modelname{}, one possible reason is that, during the lifelong learning process, models with ranking-aligned compatible leaning could effectively acquire new knowledge, and the query encoder is adjusted to better differentiate the relevant and irrelevant documents for test queries in historical sessions.
(2) Besides the forgetting issue, \modelname{}$_{\mathit{rank}}$ shows promising forward transfer ability than the models optimized with embedding alignment, probably because tight embedding alignment with existing embeddings  hinders model updating and generalizing to new queries.
(3) On the training time, our methods have significantly lower training costs compared to GSS and OCS which require gradient calculations for each training sample and MIR which requires extra estimated model updating.

\vspace{-2mm}
\section{Conclusion and Future Work}
In this work, we study a common scenario in real-world search engines, where numerous documents are continuously emerging with potential distribution drift.
To adapt the retriever to new distributions, we propose a memory-based lifelong learning method for first-stage retrieval (i.e., \modelname{}). 
By employing the selection strategy of \textit{diverse support negatives} for model updating, along with a \textit{ranking alignment objective} for backward-compatible representation learning, \modelname{} could continuously learn the retriever on unlabeled emerging documents both effectively and efficiently.
Extensive experiments on our constructed benchmarks demonstrate the superiority of \modelname{} over competitive lifelong learning baselines.

Our work presents an initial step towards solving the critical challenges in lifelong learning for first-stage retrieval.
Due to page limitations, certain promising directions remain unexplored in this study.
Firstly, it is worth investigating whether the methods proposed for domain adaptation still work well in the lifelong learning setting, as both address the distribution changes. 
Secondly, the current method does not yet have specialized techniques to handle queries related to booming topics, which presents an avenue for future research. 
In conclusion, we believe that our study, despite its limited scope, provides valuable and generalizable insights that could guide future research on this task.

\begin{acks}
This work was funded by the National Natural Science Foundation of China (NSFC) under Grants No. 61902381, the Youth Innovation Promotion Association CAS under Grants No. 2021100, the project  under Grants No. JCKY2022130C039 and 2021QY1701,  the CAS Project for Young Scientists in Basic Research under Grant No. YSBR-034, the Innovation Projectof ICT CAS under Grants No. E261090, and the Lenovo-CAS Joint Lab Youth Scientist Project. 
\end{acks}

\bibliographystyle{ACM-Reference-Format}
\bibliography{paper}

\appendix

\end{document}